# A Aplicação de uma Nova Metodologia de Ensino de Física: O Aprendizado Colaborativo

José Acácio de Barros, José Luiz Matheus Valle, Glauco S. F. da Silva, José Roberto Tagliati, Julie Remold

Quando falamos de Física, geralmente assusta a quem ouve, pois esta disciplina tem-se tornado uma grande vilã dos alunos tanto do Ensino Médio quanto do Ensino Superior. Os alunos não conseguem fazer nenhuma associação dos conceitos físicos com o seu cotidiano. Estes consideram a Física como um conjunto de fórmulas matemáticas que se deve decorar para fazer uma prova ou para aplicá-las nos exames de vestibular elaborados pelas universidades e faculdades para as quais pleiteiam uma vaga nos cursos por elas oferecidos. Para os alunos do Ensino Superior que utilizam a Física como disciplina básica em seus cursos (Engenharia, Matemática e Química) esta idéia não muda muito. Eles não vêm aplicações em suas áreas de atuação. O mesmo acontece para os alunos que entram para o curso de Física, eles não conseguem fazer uma ligação dos conceitos físicos com o dia a dia e com grande freqüência não percebem a sua importância.

Neste trabalho relatamos a experiência que tivemos quando resolvemos mudar a estrutura de ensino na disciplina Física I da Universidade Federal de Juiz de Fora (UFJF). Escolhemos esta disciplina porque é o primeiro contato com a Física que o aluno tem quando entra para algum dos cursos supra citados. Começamos a aplicação desta mudança com os alunos do curso de Física. Para estes alunos, a Física I faz parte do conjunto de disciplinas obrigatórias programadas para o primeiro semestre de curso. Atualmente este método também está sendo usado para uma turma formada por alunos dos Cursos de Matemática e Química.

O curso de Física I da UFJF envolve o estudo de mecânica newtoniana com cálculo como co-requisito (espera-se dos alunos do curso que, simultaneamente, estejam cursando também Cálculo I). Nosso curso se reúne duas vezes por semana em aulas de duas horas cada.

A principal mudança que fizemos no curso foi a utilização de métodos que estimulam a participação ativa dos alunos, utilizando *aprendizado colaborativo*[1]. Este tipo de aprendizado pode ser complementado explorando também o estudo individual que o aluno faz. Esta metodologia permite que, ao compartilharem e debaterem conceitos e idéias, os alunos dêem um salto qualitativo em seu aprendizado. Os alunos, desta maneira, colaboram tanto com seu próprio aprendizado quanto com os dos demais alunos[2]. Este dinamismo é semelhante ao da comunidade científica que debate e discute para se obter resultados.

Para implementar o aprendizado colaborativo, fizemos as seguintes modificações. As aulas de duas horas foram divididas em duas partes: uma aula expositiva e uma aula com atividades em grupo.

Na parte expositiva, utilizamos o método de Instrução por Pares (*Peer Instruction*) de Eric Mazur. Na Instrução por Pares, recitações expositivas de aproximadamente 15 minutos, apresentadas pelo professor, são seguidas de questões ou problemas conceituais relacionados ao que foi exposto (*ConcepTests*). Os Testes Conceituais tem respostas em forma de múltipla escolha. Após a apresentação dos Testes Conceituais, e após um certo tempo passar para que os alunos tenham a oportunidade de pensar sobre qual a resposta seria correta, o professor pede que toda a turma escolha uma das respostas. Neste momento, torna-se possível para o professor avaliar se a turma entendeu ou não o assunto apresentado na seção expositiva de 15 minutos. Se a maioria da turma respondeu corretamente ao Teste Conceitual, a aula segue seu curso normal. Se, por outro lado, parte da turma responde corretamente mas parte responde incorretamente, o professor instrúi os alunos a discutirem com seus vizinhos de carteira qual seria a resposta correta (por isto o nome Instrução por Pares). O processo se repete até que todos a maioria dos alunos respondam corretamente.

Além da Instrução por pares, utilizamos também mini-relatórios. Os mini-relatórios são textos de um parágrafo escritos pelos alunos após a apresentação expositiva de aproximadamente 15 minutos (similar aos Testes Conceituais). Após uma breve aula expositiva o professor orienta os alunos a responderem, num pedaço de papel, a questões como "qual foi o conceito mais confuso na aula até agora" ou "qual foi o ponto mais importante que aprendi até agora". Os mini-relatórios tem como objetivo estimular não somente a participação dos alunos em aula mas também introduzir e tentar tornar um hábito para os alunos de Física I o pensamento sobre o próprio pensamento, metacognição. Somente com este hábito os alunos podem autoavaliar e planejar seu pensamento, tornando-se capazes de aprender de maneira independente.

Na segunda parte da aula, a turma era dividia em grupos de 3 a 4 alunos onde propomos uma atividade relacionada ao assunto da aula expositiva. Estas atividades foram tiradas do *Tutorials in Introductory Physics*[3]. A ênfase dos exercícios propostos nas atividades não era na resolução de exercícios padrões, mas buscavam desenvolver os conceitos físicos importantes e a capacidade de argumentação científica. Para que se mantivesse um dinamismo entre os alunos, cada um deles tinha uma função que era diferente a cada aula. Esta função permitia que todos os alunos participassem ativamente das discussões. Eram elas então: o *Líder*, que era o responsável pela execução da atividade em questão; o *Anotador* que era responsável por registrar todas as discussões do grupo; e o *Cético*, responsável por questionar os detalhes das discussões para que tudo se esclarecesse [1]. Estes papeis tentam reproduzir em pequena escala a dialética científica, tentando fomentar a construção do conhecimento a partir dos conhecimentos individuais.

Uma componente também voltada para estimular a metacognição foi o uso de Questionários periódicos. Estes Questionários tinham em geral as seguintes três perguntas:

1) o que você aprendeu na semana passada?

2) qual foi ponto que você teve mais dificuldades na semana passada?

3) se você fosse o professor desta matéria, que pergunta faria para um aluno para descobrir se ele entendeu o assunto da semana passada?

Os Questionários não fazem parte da aula, mas sim de atividades de fora de sala de aula.

Para a participação individual na colaboração de seu próprio aprendizado preparamos listas de exercícios semanais. Estas listas contem exercícios relacionados com as aulas expositivas e com as atividades dos grupos.

Em nosso curso de Física I tivemos, surpreendentemente, poucos alunos reclamando de trabalhar em grupos. De fato, o número de alunos que reclamaram do método utilizado, incluindo as atividades em grupo, foi pequeno comparado com o número de alunos que acharam o método, como um todo, melhor do que aulas puramente expositivas. Além disso, alunos que haviam sido reprovados em semestres anteriores (até mais que uma vez) em turmas com o método tradicional se mostraram bastante entusiasmados com este método, fazendo com que os índices de evasão na disciplina tenha caído drasticamente. Vale a pena ressaltar que, curiosamente, nenhum aluno foi indiferente ao método usado, e que todos os que entrevistamos consideraram o método ou melhor ou pior do que aulas puramente expositivas, mas ninguém achou igual.

Verificamos a eficácia desta metodologia aplicando o *Inventário dos Conceitos de Força (ICF)[4]*. O ICF é um teste de questões de Mecânica em forma de Múltipla escolha onde os alunos são forçados a escolher entre as idéias do senso comum e o conceito newtoniano correto. Sua estrutura é feita baseada em temas bem definidos: Cinemática, 1°, 2º e 3° Leis de Newton, Principio da Superposição e tipos de Forças. Com isto é possível identificar o grau de compreensão dos alunos.

Para obtermos dados sobre o método em questão, aplicamos o ICF duas vezes no semestre. Um pré- teste no inicio das aulas e um pós teste no término das aulas. A vantagem da aplicação do ICF é que com os erros é possível identificar os conceitos espontâneos do senso comum. O ICF não consiste de um teste de inteligência [1].

Utilizamos como medida o ganho normalizado para o ICF. Ele é definido como a diferença entre o percentual da resposta correta no pós – teste e do pré-tese dividido pela diferença máxima entre os dois (100% menos o percentual de respostas corretas no pré – teste). O ganho é, portanto uma grandeza que varia entre 0 e 1.

Durante os semestres que utilizamos o Aprendizado Colaborativo, os resultados foram positivos, pois obtivemos um ganho normalizado superior a 0,3, e de 0,2 para as turmas onde utilizamos ensino tradicional. Estes resultados estão de acordo com a literatura internacional. Assim, concluímos que houve um aumento significativo na compreensão dos conceitos newtonianos.

**References**

1. Silva, Glauco S. F. et al. (2003). O Aprendizado Colaborativo no curso de Física I da UFJF. Artigo apresentado no XV Simpósio Nacional de Ensino de Física, 21 a 26 de março de 2003, Curitiba, PR, Brasil.
2. Collazos, C.; Guerrero, L.; Vergara, A. (2001). Aprendizaje Colaborativo: Un cambio en el rol del profesor. *Proceedings of the 3rd Workshop on Education on Computing*, Punta Arenas, Chile, November, 2001.
3. Shaffer, P. S., McDermott, L. C. & The Physics Education Group at the University of Washington (2001). *Tutorials in Introductory Physics*. Prentice Hall: New York.
4. Hestenes, D., & Halloum, E. (1992). Force Concept Inventory. *The Physics Teacher*, 30 141-158.